# An analytical performance comparison of exchanged traded funds with index funds: 2002-2010.

**Mohammad Sharifzadeh, PhD., M. Phil, CFA**

Is adjunct professor of finance, Walden University and director of research, Alpha Beta Investment Research.

**Simin Hojat, M.Phil.**

Is chief economist at Alpha Beta Investment Research.

**Abstract**  Exchanged Traded Funds (ETFs) have been gaining increasing popularity in the investment community as is evidenced by the high growth both in the number of ETFs and their net assets since 2000. As ETFs are in nature similar to index mutual funds, in this paper we examined if this growing demand for ETFs can be explained through their outperformance as compared to index mutual funds. We considered the population of all ETFs with inception dates prior to 2002 and then for each ETF found all the passive index mutual funds that had the same investment style as the selected ETF and had inception date prior to 2002. This lead to a sample of 230m paired matches for all the styles. Within each investment style we matched every ETF with all the passive index funds in that investment style and compared the performances of the matched pairs in terms of Sharp Ratios and risk adjusted buy and hold total returns for the period 200-2010. We then applied the *Wilcoxon signed rank test* to examine if ETFs had better performances than index mutual funds during the sample period. We conducted the test both within each style and for all the styles together. In terms of Sharpe Ratio, out of the 12 styles included in the sample, in 5 cases the conclusion was that ETFs outperformed index funds, for the 3 styles, U.S. Broad market, U.S large cap growth, and U.S. Reits index funds outperformed ETFs and for 4 styles there was statistically no significant difference between ETFs and index funds performances. Out of the 230 paired matches of all the styles, ETFs outperformed index mutual funds in 134 of the times in terms of Sharpe Ratio, however, the test of the hypothesis showed no statistically significant difference between ETFs and index funds performances in terms of Sharpe ratio. Break-down of the results for risk-adjusted buy and hold total returns was slightly different from those for the Sharpe Ratio, though the overall conclusion was the same. For 5 of the styles index funds outperformed ETFs, for 3 styles ETFs outperformed index funds, and for 4 styles there was statistically no significant difference between the performances of the two. Out of the 230 paired matches of all the styles, ETFs outperformed index mutual funds in 125 of the times in terms of risk adjusted buy and hold total return, however, the test of hypothesis showed no statistically significant difference between ETFs and index funds performances in terms of risk adjusted buy and hold total return. These findings indicate there is statistically no significant difference between ETFs and passive index mutual funds performances at the fund level and investors' choice between the two is related to product characteristics and tax advantages.

## 1- Introduction

Exchanged Traded Funds (ETFs) have been gaining increasing popularity in the investment community as is evidenced by their high growth both in the number of ETFs and their net assets since 2000. As ETFs and index mutual funds are generally regarded as substitutes due to the fact that they are both index-based, in this paper we will examine if the growing demand for ETFs can be explained by their outperforming index mutual funds. Most academic papers on ETF or comparisons of ETFs with index mutual funds approach the subject by discussing or examining the sources of tracking errors and tracking differences in ETFs and in index mutual funds.  Such approaches can make conclusions only about the ETFs and index mutual funds that track the same index and their findings can not be generalized to all the ETFs and index mutual funds that share the same investment style but not the same tracking index. To our knowledge there is no academic paper that compares a large sample of ETFs and index mutual funds by analyzing their risk-return characteristics over a long enough period of time to be able to make generalizations. In this research study we will identify the population of all ETFs with inception dates prior to 2002 and then for each ETF we will find all the passive index mutual funds that had the same investment style as the selected ETF and had inception date prior to 2002. Within each investment style we will match every ETF with all the passive index funds in that investment style and will compare the performances of the matched pairs in terms of Sharp Ratios and risk adjusted buy and hold total returns for the period 2000-2010.

Several authors have developed test of hypothesis for Sharpe Ratio through studying its sample statistical properties. However, these tests are all parametric which try to derive the sampling distribution of the difference between Sharpe Ratios of two portfolios from the distribution of the two portfolios' returns under certain restrictive assumptions, such as, normality, iid, lack of serial correlation, etc. They are also appropriate for large sample periods because they derive their test statistics making asymptotic assumptions and, therefore, not appropriate to use for ETFs that do not have a long history. Moreover, when there are a large number of paired-wise comparisons of Sharpe Ratios involved, like the case in this paper, the calculations and analyses applying parametric approaches become cumbersome. To overcome these conceptual and practical difficulties we will take a completely different approach for Sharpe Ratio test of hypothesis as compared to the traditional approaches; we will employ a nonparametric test of hypothesis approach, namely, the *Wilcoxon signed rank test* to examine if ETFs had better performances than index mutual funds during the sample period.

The paper is organized as the following. We start with a brief overview of ETFs; what they are, their similarities and differences with other investment companies, and their growth in comparison to all the investments companies. Then we will describe the purpose and motivation of the study followed by a literature review on ETF versus index fund debate. Next we will state our methodology and hypotheses, sampling, and results. Finally we will conclude the paper with a summary of the study and its results.

## 2- Brief overview of the ETFs

ETF is a rather new innovation in the financial industry. In the U.S. an Exchange traded fund (ETF) is a special type of registered investment company. Other types of registered investment companies are mutual funds (or open-ended funds), close ended funds, and unit investment trusts (UITs). An ETF is typically organized as a mutual fund or UIT, however, while it has some features similar to the mutual funds or UITs it has certain characteristics which make it unique and different from both mutual funds and UITs; and that is why it is categorized under a separate registered investment company.  Mutual funds are not listed with the stock exchanges and investors can only buy them from the fund company, through their financial advisor, broker-dealer, or directly.  Mutual funds take orders during Wall Street trading hours, but the transactions actually occur at the close of the market. The price the buyer pays or the seller receives is the net asset value (NAV) per share of the fund at the close of the trading day. But, an ETF can be bought or sold in the stock exchanges intra-day at market prices, which could be different from the funds instantaneous or close of the day NAV; it can be shorted, or margined just like the way these can be done with the shares of any other publicly traded company. In an open-ended fund (mutual fund) when customers buy shares from the mutual fund the fund issues new shares, cash moves in, and  the number of outstanding shares of the fund increases; and conversely when clients redeem shares cash moves out, the fund might have to sell some of the underlying securities, and  the number of fund's outstanding shares decreases. In an ETF, when investors buy or sell the ETF shares in the secondary market, there is no change in the number of ETF shares or the ETF cash versus securities balance.  However, this does not mean that the ETF is a closed-end fund or a UIT. An ETF has its own shares creation or redemption procedure.

Creation and redemption of an ETF is a unique and efficient process. ETF shares are created or redeemed through the so called *in-kind* trade process. "ETF shares are created when an "authorized participant"—typically a large institutional investor, such as a market maker or specialist—deposits the daily creation basket and/or cash with the ETF" (Investment Company Institute 2010, p. 42). In exchange, the ETF creates and delivers to the authorized participant (AP) shares of ETF, usually in blocks of 25,000 to 200,000 shares of the ETF; each block is called a *creation unit* .The AP can keep the ETF shares or sell them in the secondary market. Redemption process is just the opposite of creation process. An AP buys a large block of ETFs on the open market and delivers it to the ETF. In return the ETF delivers an equivalent basket of underlying securities to the AP. This *in-kind* exchange of underlying securities with ETF shares does not involve sale of securities by the ETF and thus does not trigger capital gains for tax purposes.

The creation and redemption feature of ETFs not only provides the needed liquidity in the ETF market but it also causes the ETFs not to deviate substantially from their net asset values (NAVs). If an ETF trades at a premium to its NAV, there is an arbitrage opportunity for the APs to buy the ETF's underlying securities, create some ETFs, and then sell them in the market bringing the ETF price towards its NAV. Conversely, if an ETF trades at a discount to its net asset value, institutional investors can purchase 25,000 to 200,000-share blocks of the ETF in the open market at the discounted price, redeem them for the underlying securities, and sell those securities at a profit. The arbitrage process will increase the market price of the ETF shares and thus closes the gap between the ETF's market price and the net asset value of the underlying portfolio. It must be noted, however, that only large investors can create or redeem ETFs; retail investors can only buy or sell the ETFs in the secondary market.

Until 2008, the U.S. Securities and Exchange Commission (the SEC) permitted only ETFs that tracked designated indexes. In early 2008, the SEC permitted creation of actively managed ETFs, which did not have to track an index. Actively managed ETFs should be transparent and must disclose the identities and weightings of the component securities and other assets held by the ETF each business day on their publicly available web sites (Investment Company Institute, 2010). In this paper we have studied ETFS with inception dates of 2001 or prior and, therefore, only passively managed ETFs are included in our sample which have been compared with their passively managed index fund counterparts.

The first ETF was created by State Street Global Advisor in partnership with the American Stock Exchange in 1993 as Standard and Poor's Depository Receipts (SPDRs). In late 1990s and early 2000s new ETFs started coming into the market and since 2002 the growth of ETF has been exponential both in terms of the number of ETFs and in terms of total net assets they represent. According to Investment Company Institute (2011), "by the end of 2010, the total number of index-based and actively managed ETFs had grown to 923, and total net assets were $992billion (page, 40)". Nonregistered ETFs, which primarily invest in commodities, currency, and futures constituted about 10% of total net assets of ETFs at year-end 2010. Figure 1 copied

from the Investment Company Institute (2011) shows the number of ETFs and their total assets since 2000.

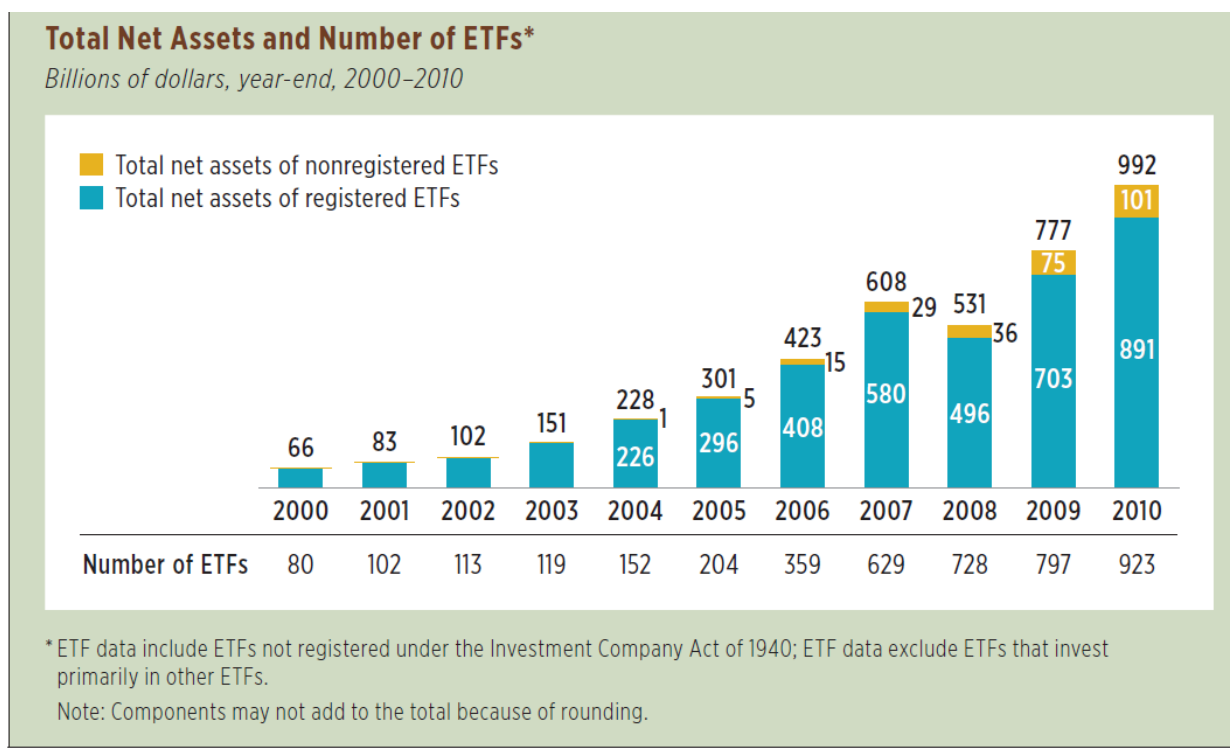

*Figure 1.* Growth of ETF 2000 to 2010 (Source: Investment Company Institute, 2011).

ETF net assets have also shown exponential growth as percentage of total

investment companies' net assets. In 1995 ETF net assets comprised only 0.03% of

total investment companies' net assets. Since then this percentage has been moving up

every year with no exception and by the end of 2010 ETF net assets represented 6.39%

of total investment companies' net assets. Table 1 shows growth of ETF net assets as

percentage of total investment companies' net assets from 1995 to 2011

Table 1

*Growth of ETF Assets as Percentage of Total Investment Companies' Net Assets*

*(Source of Data:* Investment Company Institute, 2010 and 2011).

| Year | 1995 | 1996 | 1997 | 1998 | 1999 | 2000 | 2001 | 2002 | 2003 | 2004 | 2005 | 2006 | 2007 | 2008 | 2009 | 2010 |
|---|---|---|---|---|---|---|---|---|---|---|---|---|---|---|---|---|
| Total Investment Companies' net Assets ($billions) | 3028 | 3747 | 4712 | 5791 | 7119 | 7248 | 7248 | 6680 | 7803 | 8614 | 9510 | 11167 | 12975 | 10350 | 12164 | 13100 |
| ETF net Assets ($billions) | 1 | 2 | 7 | 16 | 34 | 66 | 83 | 102 | 151 | 228 | 301 | 423 | 608 | 531 | 777 | 992 |
| ETF % of Total | 0.03% | 0.05% | 0.15% | 0.28% | 0.48% | 0.91% | 1.15% | 1.53% | 1.94% | 2.65% | 3.17% | 3.79% | 4.69% | 5.13% | 6.39% | 7.57% |

Although ETF net assets still constitute a small fraction of total investment companies' net assets (7.57% as of year-end 2010), their annual growth rate has been far above that of total investment companies' assets. ETF net assets grew from $1 billion in 1995 to $992 billion in 2010, which correspond to a total growth of 99100% for the period or an annualized geometric growth rate of 63.70% per year. For the same period total investment companies' net assets grew from $3,028 billion in 1995 to $13,100 billion in 2010, giving rise to total growth of 332.63% which corresponds to an annualized geometric growth rate of 11.03% per year. The annual growth rate of ETF net assets has been about six times more than the annual growth rate of all investment companies' net assets

## 3- Purpose and significance of this study

The purpose of this research study is to examine risk adjusted return performance of ETFs and index funds and examine if there is a statistically significant difference between the two. The findings will help to explain if investors who prefer ETF to index mutual funds are or should be doing so because of the return characteristics of ETFs, or is it because of product features of ETFs, like tax efficiency, liquidity, or tradability, that have made some investors prefer ETFs over index mutual funds. One way to accomplish this task is to calculate the tracking errors and the tracking differences for the ETFs and for the index funds and compare them. While this is a quite reasonable method to compare ETFs and index funds that replicate the same index, it can not be applied to compare all the ETFs with all the index funds and, therefore, the conclusions from such an approach can not be generalized. As we will see in the sampling section of this paper the number of ETFs that replicate the same index as an index fund is very few, particularly for the ETFs with long inception dates.

To be able to make generalization one needs to study a larger sample than those ETFs and index mutual funds that replicated the same index. In this research, to include more ETFs in the study, we compared risk- return characteristics of each ETF with risk-return characteristics of all the index funds that implemented the same investment style as that ETF. Therefore, choice of benchmark for the same investment style might prove to be one factor, amongst others, accounting for the possible performance differences between ETFs and index funds. For example for the ETF iShare [TM] IWB that implements U.S. Large Cap investment style and tracks Russell 1000 index, we compared its risk-return characteristics with all the index funds that employed U.S. Large Cap investment style, irrespective of whether they replicated Russell 1000 or another benchmark.

It should be noted that in this paper we examined risk-return characteristics of the ETFs and index funds from the fund's perspective and not from the perspective of investors in those funds. Therefore, factors like bid-ask spread and taxes on dividends or capital gains, which affect the investors' return, are not considered in this study.

## 4- Literature review on ETFs versus index funds

ETFs and index mutual funds are generally regarded as substitutes due to the fact that they are both index-based, that is, they replicate or track a specific index, by investing in the same securities and in the same proportions as the index, and, therefore, they compete for the investors in the same market. Index funds and most of the ETFs are both passively managed funds, in contrast to actively managed funds, where the fund managers engage in occasional portfolio rebalancing and market timing to enhance returns. Although there is a vast amount of literature about relative performance of index mutual funds (passive) versus active funds, there is not much academic and analytical writings in peer reviewed journals about relative performance of index funds and ETFs which are both passively managed funds, though because of their popularity they are vastly talked about in popular financial newspapers and financial blogs. Most academic papers on ETF or comparisons of ETFs with index mutual funds approach the subject by discussing or examining the sources of tracking errors and tracking differences in ETFs and in index mutual funds.  Such approaches can make conclusions only about the ETFs and index mutual funds that track the same index and their findings can not be generalized for all the ETFs and index mutual funds that share the same investment style but not the same tracking index. To our knowledge there is no academic paper that compares ETFs and index mutual funds by analyzing their risk-return characteristics over a long enough period of time to be able to make generalizations. With our approach, performance difference of ETFs and index mutual funds include not only those differences that result from the sources of tracking errors, but also accounts for their performance differences due to the choice of the tracking index for the specified investment style.

Aber, Li, and Can (2009) matched four iShare [TM] ETFs tracking S&P 500 Index, Russell 1000 growth index, MSCI small cap index, and MSCI EAFE index with four Vanguard index mutual funds tracking the same indexices. They studied daily price volatility and tracking ability of the sample ETFs as compared to their Vanguard index mutual fund counterparts between 2000 and 2006. They found that both the ETFs and the index mutual funds had approximately the same degree of correlations with their underlying indices in terms of daily returns. Furthermore, they reported based on cumulative daily returns the tracking ability of ETFs and their index mutual fund counterpart differed for different matches. For some pairs they were the same, for some pair ETFs performed better. In terms of tracking error, their study showed on average the Vanguard index mutual funds beat their corresponding iShares competitors, but by only 2 to 3 basis points.

Johnson (2008) studied 20 foreign country ETFs and the underlying index returns from 1997 to 2006 for the existence and sources of daily and monthly tracking errors. He found except for one case, market segmentation/integration of the foreign country was a significant source of tracking error. He also concluded "variables such as foreign index positive returns relative to the US index and whether the foreign exchange trades simultaneously with the US markets were significant explanatory variables in the correlation coefficients between ETFs and their underlying home index" (p. 253).

Kostovetsky (2003) studied relative performance of the ETFs and index mutual funds from the investors' point of view. He reported key areas of differences between the two to lie in management fees, shareholders transaction fees, taxation efficiency, and the qualitative factors transaction convenience, short selling, and ability to margin. His core finding was that index mutual funds are better suited for small investors and ETFs are preferable by large investors.

Rompotis (2009a) studied 20 vanguard ETFs and 12 Vanguard index mutual funds belonging to the same family in order to examine the existence and reasons for interfamily competition between ETFs and index mutual funds sponsored by the same entity. The sample period was from the funds' inceptions, which were 2004 for the ETFs and 1976 through 2004 for the index mutual funds, to the end of 2006. The study's main findings were that ETFs and index mutual funds on average present similar returns and risks, both of these investment vehicles produce lower returns than their benchmarks, they both have low tracking errors, and their tracking errors are positively correlated with expenses.

Rompotis (2009b) evaluated performance and trading characteristics of 73 iShare ETFs from October 3, 2005 to September 29, 2006 covering 250 daily observations. The finding was that iShares fail to accurately replicate the performance of their underlying indexes, they trade at a premium to their NAV, their tracking errors are strongly correlated with the expenses and risk, and that the premium was correlated positively with tracking error and negatively with the volume.

## 5- Methodology and research design

Risk adjusted portfolio performance measure was originally developed by Sharpe (1966), and further elaborated in Sharpe (1994) and Sharpe (2007). The Sharpe Ratio is a widely used metric to measure mutual funds' performance taking into account both the average return performance and the return volatility. The Sharpe Ratio, $SR_f$, for a fund $f$ is defined as :

$$SR_f = \frac{\mu_f - RFR}{\sigma_f} \qquad (1)$$

Several authors have developed test of hypothesis for Sharpe Ratio through studying its sample statistical properties. Jobnson and Korkie (1981) were the first who developed test of hypothesis for Sharpe Ratio; with subsequent correction made by Memmel (2003). They showed if returns are normally as well as independently and identically distributed (iid) then the asymptotic distribution of Sharpe Ratio (and thus distribution of the difference between two Sharpe Ratios) follows a normal distribution with its mean and variance being a function of the mean and variance of the portfolio's return. Lo (2002) derived statistical distribution of the Sharpe Ratios under different sets of assumptions using asymptotic theory. Specifically, he developed test statistics for Sharpe ratio under iid returns, stationary returns, and with time aggregation. More recently, Christie (2005) used a generalized method of moments (GMM) to derive the asymptotic distribution of Sharpe Ratio. His model does not require the iid and normality assumptions and the only restrictions of the model are stationarity and ergidicity.

However, these tests are all parametric which try to derive the sampling distribution of the difference between Sharpe Ratios of two portfolios from the distribution of the two portfolios' returns under certain restrictive assumptions, such as, normality, iid, lack of serial correlation, etc. They are also appropriate for large sample periods because they derive their test statistics making asymptotic assumptions. In this paper our Sharpe Ratios are calculated based on 2002-2010 funds' performances and, therefore, we deal with small sample period. Moreover, when there are a large number of paired-wise comparisons of Sharpe Ratios involved, like the case in this paper, the calculations and analyses applying parametric approaches become cumbersome.

To overcome above conceptual and practical difficulties we took a completely different approach for Sharpe Ratio test of hypothesis as compared to the traditional approaches; we employed a nonparametric test of hypothesis approach. Nonparametric methods do not make stringent assumptions about the population. In most cases they don't deal with population parameters and don't make assumptions about the population distribution.

There are two nonparametric methods for testing paired data, the *sign test* and the *Wilcoxon signed rank test*. Both these tests are distribution-free and don't require any population parameter. The signed test for testing paired data is stated in terms of equality of likelihood of the matched values. It only determines if values in one population are different from their matched values in the other population, or what comes to the same thing, if the differences of the matched values are positive or negative. However, the sign test does not consider the magnitude of the differences in determining the test statistic. In contrast, the *Wilcoxon signed rank test* considers both the sign of differences and the magnitude of differences (through ranking them) between the paired values and thus is a more efficient test than the signed test (Aczel, 2002). In this research we apply the *Wilcoxon signed rank test* to examine if ETFs had better performances than index mutual funds during the sample period.

We will test two sets of hypotheses with regard to performance comparisons of ETFs and index funds, one set for the traditional Sharpe Ratios and one set for the risk adjusted buy and hold total returns. We did the latter test of hypotheses because we think long-term investors might be more interested in risk adjusted buy and hold total returns than the Sharpe Ratios. The two tests do not necessarily lead to the same results, because the average return used in Sharpe Ratio calculation is the arithmetic average, derived from dividing cumulated total return by the number of years (annual returns are not compounded) whereas buy and hold total return is calculated by compounding the annual returns, which is the true measure of total return. We conducted these test of hypotheses for each style included in this study as well as for the whole paired matches of ETFs and index funds with pre 2002 inception date, paired by the investment style.

**5.1-    Definitions and terminologies**

Our analysis in this paper is based on calculating the Sharpe Ratio and risk adjusted total buy and hold return of ETFs and index mutual funds over the period 2002-2010, and then matching the results of every ETF with the results for all the index mutual funds that had the same investment style as the selected ETF. We used the style classification adopted by State Street Global Advisor, who is the sponsor of Spider (SPDR) ETFs. Not all the ETFs with pre 2002 inception dates had matching index funds. The styles for which there were matching ETFs and index funds with pre 2002 inception dates were *U.S. Broad Market, U.S. Large Cap, U.S. Mid Cap, U.S. Small Cap, U.S. Large Cap Growth, U.S. Large Cap Value, U.S. Small Cap Growth, U.S. Small Cap Value, U.S. Reits, Global, International Developed,* and *Europe.*

We are making the following notations and definitions:

$R_{et}$: Calendar-year total return for ETF *e* during year *t*, where *t= 2002,….2010*

We retrieved ETF's calendar-year total returns from the Morningstar's web site. Morningstar calculates ETF's total return by taking the change in the fund's market price during the period, reinvesting all income and capital-gains distributions during the period, and dividing by the starting market price.

$R_{it}$: Calendar-year total return for Index mutual fund *i* during year *t*, where *t= 2002,….2010*

We retrieved index mutual fund's calendar-year total returns from the Morningstar's web site. Morningstar calculates calendar-year total return of mutual funds, including index mutual funds, by taking the change in a fund's NAV, assuming the reinvestment of all income and capital gains distributions (on the actual reinvestment date used by the fund) during the period, and then dividing by the initial NAV.

$\mu_e$: Arithmetic average (mean) calendar-year total return of ETF $e$ during the period *2002-2010*

$\mu_i$: Arithmetic average (mean) calendar-year total return of index fund $i$ during the period *2002-2010*

$\sigma_{et}$: Standard deviation of ETF $e$ calendar-year total return during the period *2002-2010*

$\sigma_{it}$: Standard deviation of index fund $i$ calendar-year total return during the period *2002-2010*

$RFR$: Average annual rate of return on U.S. Government 10 year treasuries

$SR_e$: Sharp Ratio for ETF $e$ during for the period *2002-2010*

$SR_i$: Sharp Ratio for index fund $i$ for the period *2002-2010*

$SR_{diff(e,i)}^{y}$ : Difference between Sharpe Ratios of ETF fund $e$ and index fund $i$ following the same investment style $y$, that is:

$SR_{diff(e,i)}^{y}$ = $SR_{ey}$ - $SR_{iy}$ , where $e$ and $i$ have the same investment style $y$.

$BHTR_e$: Buy and hold total return of ETF $e$ for the period *2002-2010*. This is calculated using Equation (2):

$$BHTR_i = (1 + R_{e,2002})(1 + R_{e,2003})......(1 + R_{e,2010}) = \prod_{t=2002}^{2010}(1 + R_{et}) \tag{2}$$

$RABHTR_e = \dfrac{BHTR_e}{\sigma_e}$ : Risk adjusted buy and hold total return for ETF **e**

**$BHTR_i$:** Buy and hold total return of index fund **i** for the period *2002-2010*. This is calculated using Equation (3):

$$BHTR_i = (1 + R_{i,2002})(1 + R_{i,2003})......(1 + R_{i,2010}) = \prod_{t=2002}^{2010}(1 + R_{it}) \tag{3}$$

$RABHTR_i = \dfrac{BHTR_i}{\sigma_i}$ : Risk adjusted buy and hold total return for index fund **i**

$RABHTR_{dif(e,i)}^{y}$ : Difference between risk adjusted buy and hold total return of ETF fund **e** and index fund **i** following the same investment style **y,** that is:

$RABHTR_{dif(e,i)}^{y}$ = **$RABHTR_{ey}$** - **$RABHTR_{iy}$** , where **e** and **i** have the same investment style **y.**

## 5.2-  The Hypotheses

We tested two sets of hypotheses in this study, one set for comparing the Sharpe Ratios of ETFs with the Sharpe ratios of their paired index funds, and one set for comparing risk adjusted buy and hold total returns of ETFs with risk adjusted buy and hold total returns of their paired index funds. The criterion for paring was the investment style, every ETF was paired with all the index funds that followed the same investment style as that ETF.

**Hypothesis 1**: ETFs have higher Sharpe Ratios than index mutual funds. Symbolically this is expressed as the common null (**H$_0$**) and alternate hypotheses (**H$_1$**):

**H$_0$**: $SR^y_{diff(e,i)} \leq 0$

**H$_1$** : $SR^y_{diff(e,i)} > 0$

Where, $SR^y_{diff(e,i)}$ is the difference between Sharpe Ratios of ETF fund **e** and index fund **i** following the same investment style **y** or the period 2002-2010. We tested hypothesis 1 for the paired ETFs and index funds for each style separately as well as over all the styles.

**Hypothesis 2**: ETFs have higher risk adjusted buy and hold total return than index mutual funds. Symbolically this is expressed as :

**H$_0$**: $RABHTR^y_{dif(e,i)} \leq 0$

**H$_1$** : $RABHTR^y_{dif(e,i)} > 0$

Where, $RABHTR^y_{dif(e,i)}$ is the difference between risk adjusted buy and hold total return of ETF fund **e** and index fund **i** following the same investment style **y** for the period 2002-2010. We tested hypothesis 2 for the paired ETFs and index funds for each style separately as well as over all the styles.

For both hypotheses 1 and 2 we used the *Wilcoxon signed rank test* at 5% significance level. The test statistic, T, for *Wilcoxon signed rank test* is defined as the smaller of sums of positive ranks and sums of negative ranks, that is:

$$T = min\{ \textstyle\sum( + ), \sum( - )\} \qquad (4)$$

For large samples, the distribution of the *Wilcoxon* statistic tends to a normal probability distribution with mean, $\mu_T$, and standard deviation, $\sigma_T$, given by:

$$\mu_T = \frac{n( n+1 )}{4}$$
$$\sigma_T = \sqrt{\frac{n( n+1 )( 2n+1 )}{24}}$$

Where, **n** is the number of paired observations

Therefore, for large samples (**n**>25) the z test can be used to conduct the *Wilcoxon signed rank*

## 5.3- Sampling

To assess performance of a financial asset one needs to consider a long history of the asset's performance. ETF's do not have a long history. Currently, as of January 17, 2011 there were 975 U.S. listed ETFs in the market (http://www.masterdata.com). However, most of these ETFs were created after 2005. The first ETF was introduced in 1999 and until the year 2000 there were not many ETFs in the market. The years 2000 and 2001were the first years in which many ETFs were created. Therefore, in order to have a relatively large sample of ETFs with long history we limited our analysis to all the ETFs that were created in 2001 and prior, and obtained their annual return performances for the fiscal years 2002-2010. To get the list of pre 2002 ETFs, we first downloaded list of all ETFs from State Street Global Advisor's web site and then picked all the ETFs with inception dates being 2001 or before. This gave us 100 ETFs, mostly iShares and Spiders. We then grouped these ETFs according to the investment style categorization State Street Global Advisor, who is the sponsor of Spider (SPDR) ETFs. Next, we retrieved the list of passive index mutual funds with pre 2002 inception date from the IndexUnivers web site (http://www.indexuniverse.com/ecs) and from vanguard web site (http://www.vanguard.com). We then grouped the passive index mutual funds using the style categorization mentioned above. We eliminated the enhanced index funds from the list, as these funds have some degree of management in the form of computerized stock selection or else. Next we paired  every ETF with all the index mutual funds that had the same investment style as the selected ETF. Not all the ETFs with pre 2002 inception dates had matching index funds. For example for U.S. Consumer Discretionary ETFs, there were IYC (ishares) and XYL (SPDR) with pre 2002 inception date, but there were no index fund with pre 2002 inception date to be matched with  those ETFs. The styles for which there were matching ETFs and index funds with pre 2002 inception dates were *U.S. Broad Market, U.S. Large Cap, U.S. Mid Cap, U.S. Small Cap, U.S. Large Cap Growth, U.S. Large Cap Value, U.S. Small Cap Growth, U.S. Small Cap Value, U.S. Reits, Global, International Developed,* and *Europe*. Out of the 100 ETFs with pre 2002 inception date 67 of them did not have a matching index mutual fund to be paired with and only 34 of them had matching index mutual funds. We identified 66 passive index mutual funds with pre 2002 inception dates to match with our selected ETF sample. Completing the process of pairing ETFs with the same style index funds having inception dates prior to 2002, we came up with 230 paired matches.  Table 3 below shows the number of ETFs, index funds, and paired matches for each style and for all the styles.

Table 3

*Number of pre 2002 inception date ETFs, index mutual funds, and paired matches for each investment style and all the styles.*

| Investment style | Number of ETFs (Pre 2002 Inception) | Number of index funds | Number of paired matches |
|---|---|---|---|
| U.S. Broad Market | 4 | 7 | 28 |
| U.S. Large Cap | 6 | 15 | 90 |
| U.S. Mid Cap | 3 | 20 | 60 |
| U.S. Small Cap | 2 | 7 | 14 |
| U.S. Large Cap Growth | 3 | 2 | 6 |
| U.S. Large Cap Value | 3 | 2 | 6 |
| U.S. Small Cap Growth | 3 | 2 | 6 |
| U.S. Small Cap Value | 3 | 2 | 6 |
| U.S. Reits | 3 | 2 | 6 |
| Global | 2 | 1 | 2 |
| International Developed | 1 | 3 | 3 |
| Europe | 1 | 3 | 3 |
| **All Styles** | **34** | **66** | **230** |

## 6- Results

We reject the null hypothesis, and accept the alternate hypothesis, when the sample *p* value is less than or equal to 5% and we reject the alternate hypothesis when the sample *p* value greater than 5%. Rejection of the null hypothesis will lead to acceptance of the alternate hypothesis implying that ETFs outperformed index funds in terms of Sharpe Ratio or risk-adjusted buy and hold total returns for the sample period. However, rejection of the alternate hypothesis does not imply that index funds outperformed ETFs; rather, It implies that either there was statistically no significant difference between performances of the two or index funds outperformed ETFs depending the magnitude of the *p*-value. In such a case, where the *p* value greater than 5% we can conclude that index funds outperformed ETFs should the *p* value be greater than 95%, otherwise the conclusion is that there is no significant difference between performances of the two.

**6.1- Results for Sharp Ratio**

Out of the 12 styles included in the sample, in 5 cases the conclusion was that ETFs outperformed index funds in terms of Sharpe ratio. For the 3 styles, U.S. Broad market, U.S large cap growth, and U.S. Reits, index funds outperformed ETFs and for 4 styles there was statistically no significant difference between ETFs and index funds performances. Out of the 230 paired matches of all the styles, ETFs outperformed index mutual funds in 134 of the times in terms of Sharpe Ratio, however, the test of the hypothesis showed no statistically significant difference between ETFs and index funds performances in terms of Sharpe ratio. Summary of the test results for Sharpe ratio is provided in Table 4

Table 4

*Summary of the Test of Hypotheses Results for Sharpe Ratio*

| Investment style | Number of ETFs (Pre 2002 Inception) | Number of index funds | Number of paired matches | Number of times ETFs outperformed index funds | Test of Hypothesis Results (at 5%) |
|---|---|---|---|---|---|
| U.S. Broad Market | 4 | 7 | 28 | 9 | Index funds outperformed ETFs |
| U.S. Large Cap | 6 | 15 | 90 | 64 | ETFs outperformed Index funds |
| U.S. Mid Cap | 3 | 20 | 60 | 45 | ETFs outperformed Index funds |
| U.S. Small Cap | 2 | 7 | 14 | 6 | No significant difference |
| U.S. Large Cap Growth | 3 | 2 | 6 | 0 | Index funds outperformed ETFs |
| U.S. Large Cap Value | 3 | 2 | 6 | 2 | No significant difference |
| U.S. Small Cap Growth | 3 | 2 | 6 | 2 | No significant difference |
| U.S. Small Cap Value | 3 | 2 | 6 | 6 | ETFs outperformed Index funds |
| U.S. Reits | 3 | 2 | 6 | 0 | Index funds outperformed ETFs |
| Global | 2 | 1 | 2 | 0 | Index funds outperformed ETFs |
| International Developed | 1 | 3 | 3 | 0 | No significant difference |
| Europe | 1 | 3 | 3 | 0 | No significant difference |
| All Styles | 34 | 66 | 230 | 134 | No significant difference |

## 6.2- Results for risk-adjusted buy and hold total returns

Break-down of the results for risk-adjusted buy and hold total returns was slightly different from those for the Sharpe Ratio, though the overall conclusion was the same. For 5 of the styles index funds outperformed ETFs, for 3 styles ETFs outperformed index funds, and for 4 styles there was statistically no significant difference between the performances of the two. Out of the 230 paired matches of all the styles, ETFs outperformed index mutual funds in 125 of the times in terms of risk adjusted buy and hold total returns, however, the test of hypothesis showed no statistically significant difference between ETFs and index funds performances in terms of risk adjusted buy and hold total returns. Summary of the test results for risk-adjusted buy and hold total returns is provided in Table 5

Table 5

*Summary of the Test of Hypotheses Results for* risk-adjusted buy and hold total returns.

| Investment style | Number of ETFs (Pre 2002 Inception) | Number of index funds | Number of paired matches | Number of times ETFs outperforming index funds | Test of Hypothesis Results (at 5%) |
|---|---|---|---|---|---|
| U.S. Broad Market | 4 | 7 | 28 | 8 | Index funds outperformed ETFs |
| U.S. Large Cap | 6 | 15 | 90 | 62 | ETFs outperformed Index funds |
| U.S. Mid Cap | 3 | 20 | 60 | 42 | ETFs outperformed Index funds |
| U.S. Small Cap | 2 | 7 | 14 | 3 | Index funds outperformed ETFs |
| U.S. Large Cap Growth | 3 | 2 | 6 | 0 | Index funds outperformed ETFs |
| U.S. Large Cap Value | 3 | 2 | 6 | 2 | No significant difference |
| U.S. Small Cap Growth | 3 | 2 | 6 | 2 | No significant difference |
| U.S. Small Cap Value | 3 | 2 | 6 | 6 | ETFs outperformed Index funds |
| U.S. Reits | 3 | 2 | 6 | 0 | Index funds outperformed ETFs |
| Global | 2 | 1 | 2 | 0 | Index funds outperformed ETFs |
| International Developed | 1 | 3 | 3 | 0 | No significant difference |
| Europe | 1 | 3 | 3 | 0 | No significant difference |
| All Styles | 34 | 66 | 230 | 125 | No significant difference |

# 7- Summary and conclusions

In this paper we matched ETFs and index funds based on the investment styles and compared their paired-wise performances. We proposed two hypotheses; one for comparing the Sharpe ratios and one for comparing the buy and hold total returns of ETFs with index mutual funds over the sample period 2002-2010 using the nonparametric test of *Wilcoxon signed rank test*. The results indicated that although over 50% of the selected ETFs outperformed their index fund counterparts but this is not statistically significant and the sample evidence does not support the proposition that ETFs outperform index funds. This suggests investors' choice between ETFs and index mutual funds, as two competing investment products, depends more on product features than on the performances at the funds' level.

Some limitations of this study which could be the subject of further research include (a) the study covers the period 2002-2010, results may not be true for post 2002 inception dates, (b) the sample does not include index funds with pre 2002 inception dates, only index funds that had a matching style with an ETF were included in the sample, and (c) choice of other matching criteria could lead to different results.

**List of References**

Aber, J. W., Li, D., and Can, L. (2009). Price volatility and tracking ability of ETFs. *Journal of Asset Management*, *10*, 210-221.

Aczel. A. D. (2002). *Complete business statistics.* New York: McGraw-Hill Irwin.

Christie, S. (2005). Is the Sharpe Ratio useful in asset allocation? *Applied Finance Center, Macquarie University, MAFC Research Papers No. 31,* 1-51.

Gastineau, G., L. (2004, Winter). The benchmark index ETF performance problem: A simple solution. *The Journal of Portfolio Management,* 96-103.

Investment Company Institute (2010). Investment company fact book: *A review of trends and activity in the investment company industry. Retrieved from* www.icifactbook.org*, May 17, 2011*

Investment Company Institute (2011). Investment company fact book: *A review of trends and activity in the investment company industry. Retrieved from* www.icifactbook.org*, May 17, 2011*

Jobson, D. and Korkie, B. (1981). Performance hypothesis testing with the Sharpe and Treynor measures. *Journal of Finance, 36*(4), 889-908.

Johnson, W. F. (2008). Tracking errors of exchange traded funds. *Journal of Asset Management, 10*(4), 253-262.

Jobson, J.D.and Korkie, B. (1981). Performance hypothesis testing with the Sharpe and Treynor measures. *Journal of Finance 36*, 889-908.

Kostovetsky, L. (2003, Summer). Index mutual funds and Exchange-traded funds: A comparison of two methods of passive investment. *The Journal of Portfolio Management*,

Lo, A., W. (2002). The Statistics of Sharpe Ratios. *Financial Analyst Journal 58*(4), 36-52.

Memmel, C. (2003). Performance hypothesis testing with the Sharpe ratio. *Finance Letters*, *1* (1), 21-23.

Rompotis, G., G. (2009a). Interfamily competition on index tracking: The case of the vanguard ETFs and index funds. *Journal of Asset Management*, *10*(4), 263-278

Rompotis, G., G. (2009b). Performance and trading characteristics of iShares: An evaluation. *The Icfa Journal of Applied Finance*, *15*(7), 24-39.

Sharpe, W. F. (1966). Mutual fund performance part 2, supplement on security prices. *Journal of Business*, *31*(1), 119-138.

Sharpe, W. F. (1994). The Sharpe ratio. *Journal of Portfolio Management*, *21*(1), 49-59

Sharpe, W. F. (2007). The expected utility asset allocation. *Financial Analyst Journal*, *93*(5), 18-30

Appendix A

List of TEFs and matched Index Funds Studied